\documentstyle[12pt,aps,epsf]{revtex}
\begin{document}
\draft
\title{Uncertainties of predictions in models of eternal inflation}
\author{Serge Winitzki and Alexander Vilenkin}
\date{\today }
\address{Institute of Cosmology, Department of Physics and Astronomy,\\
Tufts University, Medford, MA 02155, USA}
\maketitle

\begin{abstract}
In a previous paper \cite{MakingPredictions}, a method of comparing the
volumes of thermalized regions in eternally inflating universe was
introduced. In this paper, we investigate the dependence of the results
obtained through that method on the choice of the time variable and factor
ordering in the diffusion equation that describes the evolution of eternally
inflating universes. It is shown, both analytically and numerically, that
the variation of the results due to factor ordering ambiguity inherent in
the model is of the same order as their variation due to the choice of the
time variable. Therefore, the results are, within their accuracy, free of
the spurious dependence on the time parametrization.
\end{abstract}


\section{Introduction}

The parameters we call ``constants of Nature'' can take different values in
different parts of the universe and in different disconnected universes in
the ensemble described by the cosmological wave function \cite{Coleman,Linde}%
. The probability distribution for the constants can be determined with the
aid of the ``principle of mediocrity'' \cite{Predictions,Mediocrity}, which
asserts that we are ``typical'' among the civilizations inhabiting this
ensemble. The resulting probabilities are then proportional to the physical
volumes occupied by thermalized regions with given values of the constants 
\cite{Constants}, and the preferred values tend to be the ones that give the
largest amount of inflation.

This prescription encounters a difficulty when applied to models where
inflation is ``eternal''. In such models, the universe consists of a number
of isolated thermalized regions embedded in a still-inflating background.
New thermalized regions are constantly being formed, but the inflating
domains that separate them expand so fast that the universe never
thermalizes completely \cite{EternalInflation1,EternalInflation2}. In an
eternally inflating universe, the thermalized volumes ${\cal V}_{*}^{(j)}$
diverge as $t\rightarrow \infty $. (Here, the index $j$ labels different
types of thermalized regions which have different values of the constants of
Nature.) If one simply introduces a time cutoff by including only parts of
the volumes that thermalized prior to some moment of time $t_c$, then one
finds that the ratio 
\[
r=\lim_{t_c\rightarrow \infty }\frac{{\cal V}_{*}^{\left( 1\right) }}{{\cal V%
}_{*}^{\left( 2\right) }} 
\]
is extremely sensitive to the choice of the time coordinate $t$. For
example, cutoffs at a fixed proper time and at a fixed scale factor give
drastically different results \cite{LLMBig}. An alternative procedure \cite
{MakingPredictions} is to introduce a cutoff at the time $t_c^{(j)}$ when
all but a small fraction $\varepsilon $ of the co-moving volume destined to
thermalize into regions of type $j$ has thermalized. The value of $%
\varepsilon $ is taken to be the same for all types of thermalized regions,
and for all universes in the ensemble, but the corresponding cutoff times $%
t_c^{(j)}$ are generally different. The limit $\varepsilon \rightarrow 0$ is
taken after calculating the probability distribution for the constants. It
was shown in \cite{MakingPredictions} that the resulting probabilities are
rather insensitive to the choice of time parametrization.

Although this can certainly be regarded as progress, the situation is still
not completely satisfactory. What do we make of the residual, ``mild''
dependence of the probabilities on the choice of the time coordinate $t$?
This dependence is particularly worrisome when one tries to compare
probabilities for different universes in the quantum-cosmological ensemble.
Using the $\varepsilon $-procedure, one finds \cite{MakingPredictions} that
the probability distribution has an infinitely sharp peak at the highest
value of the ratio $\tilde \gamma /\gamma $, where $-\tilde \gamma $ and $%
\gamma $ are the smallest eigenvalues of the diffusion equation for the
probability distribution of the inflaton field $\varphi $, with
normalization to the physical and coordinate volume, respectively. This
sharp prediction does not mix well even with a ``mild'' ambiguity in the
value of $\tilde \gamma /\gamma $.

An additional source of ambiguity in predictions of eternal inflation is in
the form of the diffusion equation itself. The equation was introduced in
Refs. \cite{EternalInflation1,LLMBig,Starobinsky,GLindeM,Mijic,Japanese} to
describe the effect of quantum fluctuations on the evolution of the inflaton
field. It gives accurate results for sufficiently flat inflaton potentials $%
V\left( \varphi \right) $, provided that the magnitude of $V\left( \varphi
\right) $ is well below the Planck energy density, $V\left( \varphi \right)
\ll 1$. [Here and below we use Planck units in which $\hbar =c=G=1$.] By its
nature, however, the equation is an approximation to quantum field theory in
a curved spacetime, or, if gravity is to be adequately included, to the
Wheeler-DeWitt equation for the cosmological wave function. This approximate
nature of the diffusion equation manifests itself, in particular, in the
ambiguity in the ordering of factors $\partial /\partial \varphi $ and $%
D\left( \varphi \right) $, where $D\left( \varphi \right) $ is the
field-dependent diffusion coefficient.

In the present paper we shall analyze the uncertainties in predictions of
the principle of mediocrity resulting both from the choice of the time
parameter and of the factor-ordering in the diffusion equation. It will be
shown that the uncertainties of the probabilities due to these two
ambiguities are of the same order of magnitude. Since the factor-ordering
ambiguity is inherent in the diffusion approximation, the corresponding
uncertainty can be regarded as the bound on the accuracy allowed by the
model. Our result is then that, within this accuracy, the probabilities are
independent of time parametrization. This is a hopeful sign, since it
suggests that a complete independence of time parametrization can be
achieved in a more fundamental approach.

The paper is organized as follows. In the next section, we introduce the
necessary formalism. Sections \ref{ANALYTIC} and \ref{NUMERIC} present our
analytic and numerical results. Conclusions and a discussion of some open
issues follow in section \ref{DISCUSSION}.

\section{The diffusion equation}

\begin{figure}[tbp]
\epsfysize 5 cm \epsffile{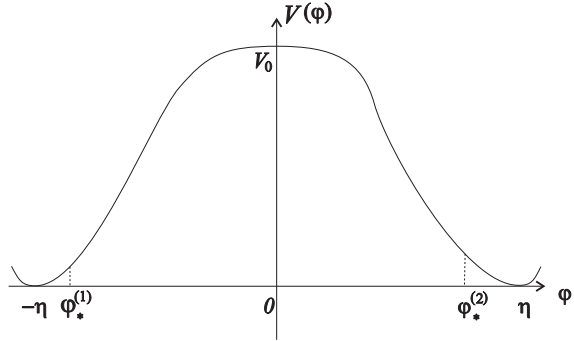}
\caption{Inflaton potential for a new inflationary scenario.}
\label{InflatonPotential}
\end{figure}

We shall consider ``new'' inflation with a potential $V\left( \varphi
\right) $ of the form illustrated in Fig.~\ref{InflatonPotential}. The
potential has two minima and the values $\varphi _{*}^{\left( 1\right) }$
and $\varphi _{*}^{\left( 2\right) }$ near the minima correspond to the end
of inflation. We will be interested in the relative probability ${\cal P}%
^{\left( 2\right) }/{\cal P}^{\left( 1\right) }$ for the two minima. In this
section, we shall first introduce the parameters $\alpha $ and $\beta $,
representing the freedom of time parametrization and factor ordering. We
shall then bring the diffusion equation to a convenient self-adjoint form,
and finally derive a general expression for~${\cal P}^{\left( 2\right) }/ 
{\cal P}^{\left( 1\right) }$.

\subsection{Parametrization of choices of time and factor ordering}

The evolution of the inflaton field $\varphi $ during inflation can be
described by a probability distribution $P\left( \varphi ,t\right)d\varphi $
which is interpreted, up to a normalization, as the (either co-moving or
physical) volume of regions with a particular value of $\varphi $ in the
interval $d\varphi$. We shall use the notation $P\left( \varphi ,t\right) $
for the co-moving and $\tilde P\left( \varphi ,t\right) $ for the physical
volume probability. The latter satisfies the diffusion equation which can be
written as

\begin{equation}
\partial _t\tilde P=\partial _\varphi \left[ D\left( \varphi \right)
^{1/2+\beta }\partial _\varphi \left( D\left( \varphi \right) ^{1/2-\beta }%
\tilde P\right) -v\left( \varphi \right) \tilde P\right] +3H\left( \varphi
\right) ^\alpha \tilde P  \label{TdFullEquPp}
\end{equation}
where 
\[
H\left( \varphi \right) =\sqrt{\frac{8\pi }3V\left( \varphi \right) } 
\]
is the Hubble constant, $V\left( \varphi \right) $ is the inflaton
potential, $D\left( \varphi \right) $ is the diffusion coefficient, 
\begin{equation}
D\left( \varphi \right) =\frac 1{8\pi ^2}H\left( \varphi \right) ^{\alpha
+2},  \label{DiffCoef}
\end{equation}
and 
\begin{equation}
v\left( \varphi \right) =-\frac 1{4\pi }H\left( \varphi \right) ^{\alpha
-1}\partial _\varphi H\left( \varphi \right)  \label{DriftVel}
\end{equation}
is the average ``drift velocity'' corresponding to the slow roll. Eq. (\ref
{TdFullEquPp}) is accurate provided that the condition of slow roll, 
\begin{equation}
\left| H^{\prime }\left( \varphi \right) \right| \ll 2\pi H\left( \varphi
\right) ,  \label{SlowRoll}
\end{equation}
is satisfied.

The initial distribution $\tilde P\left( \varphi ,t=0\right) $ can be
derived from the cosmological wave function of the nucleated universe \cite
{QuantumCosmology}. For the ``tunneling'' wave function, 
\begin{equation}
\tilde P\left( \varphi ,t=0\right) \propto \exp \left( -\frac 3{8V\left(
\varphi \right) }\right) .  \label{TunWF}
\end{equation}
\ Essentially the same results are obtained by taking any bell-shaped
initial distribution peaked around the maximum of the inflaton potential $%
V\left( \varphi \right) $.

The solution is subject to the following boundary conditions at
thermalization points $\varphi =\varphi _{*}^{\left( 1,2\right) }$: 
\begin{equation}
D\left( \varphi \right) ^{1/2+\beta }\partial _\varphi \left( D\left(
\varphi \right) ^{1/2-\beta }P\right) =0.  \label{NoDiffBC}
\end{equation}
The boundary condition (\ref{NoDiffBC}) means that diffusion is constrained
to vanish at points $\varphi _{*}^{\left( 1,2\right) }$ which are found from
the ``thermalization condition'', 
\begin{equation}
\left| H^{\prime }\left( \varphi _{*}^{\left( 1,2\right) }\right) \right|
\simeq 2\pi H\left( \varphi _{*}^{\left( 1,2\right) }\right) .
\label{ThermCond}
\end{equation}
This equality signifies the breakdown of the slow roll condition (\ref
{SlowRoll}). Since (\ref{ThermCond}) is an order-of-magnitude relation, the
exact values of $\varphi _{*}^{\left( 1,2\right) }$ depend on the choice of
the constant of order $1$ in (\ref{ThermCond}). Although this introduces an
ambiguity in our model, we note that diffusion, which represents quantum
fluctuations, is already negligibly small in the regions dominated by the
slow roll, and the ambiguity in the choice of points $\varphi _{*}^{\left(
1,2\right) }$ at which diffusion is constrained to exactly vanish does not
significantly influence the solution of the diffusion equation \cite
{AmbigPhi}. The ``no-diffusion'' boundary condition (\ref{NoDiffBC}) was
introduced in \cite{LLMBig}.

In equation (\ref{TdFullEquPp}), different choices of time variable and
factor ordering are represented by choices of the parameters $\alpha $ and $%
\beta $. The parameter $\alpha $ is equal to the physical dimension of the
time variable $t$, which is related to the proper time $\tau $ by 
\begin{equation}
dt=H\left( \varphi \right) ^{1-\alpha }d\tau  \label{TimeRel}
\end{equation}
(so that $\alpha =1$ corresponds to the proper time parametrization, $t=\tau 
$, and $\alpha =0$ corresponds to using the scale factor as a time
variable). The choice of the factor ordering is parametrized by $\beta $
which is defined so that $\beta =0$ gives the so called Stratonovich factor
ordering \cite{LLMBig,Starobinsky}. A similar parametrization of factor
ordering in the diffusion equation (\ref{TdFullEquPp}) was used by Salopek
and Bond \cite{Salopek}.

The equation for the coordinate-volume distribution $P\left( \varphi
,t\right) $ is identical to (\ref{TdFullEquPp}), except for the absence of
the expansion term: 
\begin{equation}
\partial _tP=\partial _\varphi \left[ D\left( \varphi \right) ^{1/2+\beta
}\partial _\varphi \left( D\left( \varphi \right) ^{1/2-\beta }P\right)
-v\left( \varphi \right) P\right]  \label{TdFullEquPc}
\end{equation}

More generally, one could consider a time variable related to the proper
time $\tau $ by 
\begin{equation}
dt=T\left( \varphi \right) d\tau ,  \label{GenTime}
\end{equation}
with $T\left( \varphi \right) $ an arbitrary function; the choice (\ref
{TimeRel}) corresponds to $T\left( \varphi \right) =H\left( \varphi \right)
^{1-\alpha }$. With a general time function, the diffusion coefficient would
change to 
\begin{equation}
D\left( \varphi \right) =\frac{H\left( \varphi \right) ^3}{8\pi ^2T\left(
\varphi \right) }
\end{equation}
and the drift velocity would become 
\begin{equation}
v\left( \varphi \right) =-\frac{H^{\prime }}{4\pi T\left( \varphi \right) }.
\end{equation}

Also, the factor ordering in Eqs. (\ref{TdFullEquPp}),(\ref{TdFullEquPc})
could be generalized to insertion of an arbitrary function $h\left( \varphi
\right) $ which would change the diffusion term in those equations to 
\begin{equation}
D\left( \varphi \right) ^{1/2+\beta }\partial _\varphi \left( D\left(
\varphi \right) ^{1/2-\beta }P\right) \rightarrow D\left( \varphi \right)
^{1/2}h\left( \varphi \right) \partial _\varphi \left( D\left( \varphi
\right) ^{1/2}\frac 1{h\left( \varphi \right) }P\right) .  \label{GenFactor}
\end{equation}
The factor ordering of Eqs. (\ref{TdFullEquPp}),(\ref{TdFullEquPc})
corresponds to the choice $h\left( \varphi \right) =D\left( \varphi \right)
^\beta $.

We will not consider these possibilities here, since the physical reasons
for the ``correct'' choice of functions $T\left( \varphi \right) $ and $%
h\left( \varphi \right) $ are not clear. In the following discussion, we
shall use exclusively the parameters $\alpha $ and $\beta $ to explore the
ambiguities related to these choices.

\subsection{Self-adjoint form of the diffusion equation}

The late-time behavior of the distribution functions described by Eqs. (\ref
{TdFullEquPp}),(\ref{TdFullEquPc}) is determined by the eigenfunctions 
\begin{equation}
P\left( \varphi ,t\right) =e^{-\gamma t}\psi ^{\left( 1\right) }\left(
\varphi \right) ,\quad \tilde P\left( \varphi ,t\right) =e^{\tilde \gamma t}%
\tilde \psi ^{\left( 1\right) }\left( \varphi \right) ,  \label{EigenFun}
\end{equation}
having highest eigenvalues $\tilde \gamma $ and $-\gamma $. The equations
for $\tilde \psi \left( \varphi \right) $ and $\psi \left( \varphi \right) $
are stationary forms of (\ref{TdFullEquPp}),(\ref{TdFullEquPc}), 
\begin{eqnarray}
\partial _\varphi \left[ D\left( \varphi \right) ^{1/2+\beta }\partial
_\varphi \left( D\left( \varphi \right) ^{1/2-\beta }\tilde P\right)
-v\left( \varphi \right) \tilde P\right] +3H\left( \varphi \right) ^\alpha 
\tilde P &=&\tilde \gamma \tilde P,  \label{FullEquPp} \\
\partial _\varphi \left[ D\left( \varphi \right) ^{1/2+\beta }\partial
_\varphi \left( D\left( \varphi \right) ^{1/2-\beta }P\right) -v\left(
\varphi \right) P\right] &=&-\gamma P.  \label{FullEquPc}
\end{eqnarray}
The sign convention in the definition of $\gamma $ and $\tilde \gamma $
reflects the fact that the highest value of $\tilde \gamma $ and the
smallest value of $\gamma $ are both positive. (If the highest value of $%
\tilde \gamma $ is negative, then there is no eternal inflation. See also
section \ref{EstimateGammas} below.)

Equations (\ref{FullEquPp}),(\ref{FullEquPc}) can be transformed to a
manifestly self-adjoint form \cite{Method}. One introduces a new independent
variable $z$, 
\begin{equation}
z=\int \left[ 2D\left( \varphi \right) \right] ^{-1/2}d\varphi ,\quad
P\left( \varphi \right) =\left[ 2D\left( \varphi \right) \right]
^{-1/2}f\left( z\right) ,  \label{ChangeVar}
\end{equation}
and a further substitution 
\begin{equation}
f\left( z,t\right) =g\left( z,t\right) D\left( z\right) ^{\beta /2}\exp
\left( \frac \pi {2H\left( z\right) ^2}\right) \equiv g\left( z,t\right)
\exp \left( \int b\left( z\right) dz\right) ,
\end{equation}
where we have defined 
\begin{equation}
b\left( z\right) =-\pi \frac{H_z^{\prime }}{H^3}+\frac \beta 2\frac{%
D_z^{\prime }}D.  \label{EffPotential}
\end{equation}
Equation (\ref{TdFullEquPc}) then leads to the following equation for $%
g\left( z,t\right) $: 
\begin{equation}
\partial _tg=\frac 12\partial _{zz}g+(-\frac 12b_z^{\prime }-\frac 12b^2)g.
\label{TdSchrEquPc}
\end{equation}
This form of the equation was derived in \cite{Mijic} for the case of $%
\alpha =\beta =0$.

Under the same transformations, equation (\ref{TdFullEquPp}) for the
physical-volume distribution becomes 
\begin{equation}
\partial _t\tilde g=\frac 12\partial _{zz}\tilde g+\left( -\frac 12%
b_z^{\prime }-\frac 12b^2+3H^\alpha \right) \tilde g,  \label{TdSchrEquPp}
\end{equation}
The transformed versions of Eqs. (\ref{FullEquPp}),(\ref{FullEquPc}) can be
written as: 
\begin{eqnarray}
-\frac 12\tilde g^{\prime \prime }+\left( \frac 12b_z^{\prime }+\frac 12%
b^2-3H^\alpha \right) \tilde g &=&-\tilde \gamma \tilde g,  \label{SchrEquPp}
\\
-\frac 12g^{\prime \prime }+\left( \frac 12b_z^{\prime }+\frac 12b^2\right)
g &=&\gamma g,  \label{SchrEquPc}
\end{eqnarray}
where primes denote differentiation with respect to $z$. The boundary
conditions (\ref{NoDiffBC}) become 
\begin{equation}
\partial _z\left[ D^{-\beta /2}\exp \left( \frac \pi {2H^2}\right) g\left(
z\right) \right] =0,  \label{zBC}
\end{equation}
and they are to be imposed at points $z_{*}^{\left( 1,2\right) }$
corresponding to the thermalization points $\varphi _{*}^{\left( 1,2\right)
} $ of (\ref{ThermCond}).

Equations (\ref{SchrEquPp}),(\ref{SchrEquPc}) have the form of a stationary
Schr\"odinger equation for a one-dimensional motion in a potential $U\left(
z\right) $, 
\begin{equation}
-\frac 12g^{\prime \prime }+U\left( z\right) g=Eg,  \label{SchrEqu}
\end{equation}
with 
\begin{equation}
\tilde U\left( z\right) =\frac 12b^{\prime }+\frac 12b^2-3H^\alpha
\label{UzPp}
\end{equation}
for the physical-volume distribution and 
\begin{equation}
U\left( z\right) =\frac 12b^{\prime }+\frac 12b^2  \label{UzPc}
\end{equation}
for the coordinate-volume distribution. With boundary conditions (\ref{zBC}%
), the operator $\left( -1/2\right) \partial _{zz}$ appearing in the left
hand side of (\ref{SchrEqu}) is non-negative if 
\begin{equation}
\int\limits_{z_{*}^{\left( 1\right) }}^{z_{*}^{\left( 2\right) }}g\left(
z\right) \left( -\frac 12\partial _{zz}g\right) dz=\left[ -g\left( z\right)
^2\frac{H_z^{\prime }}{4H^3}\left( 2\pi +H^2\beta \left( \alpha +2\right)
\right) \right] _{z_{*}^{\left( 1\right) }}^{z_{*}^{\left( 2\right) }}+\frac 
12\int\limits_{z_{*}^{\left( 1\right) }}^{z_{*}^{\left( 2\right) }}\left[
g^{\prime }\left( z\right) \right] ^2dz\geq 0
\end{equation}
for any function $g\left( z\right) $. This holds for a potential of Fig. \ref
{InflatonPotential} since $V^{\prime }\left( \varphi _{*}^{\left( 2\right)
}\right) <0$ and $V^{\prime }\left( \varphi _{*}^{\left( 1\right) }\right)
>0 $ and so $H_z^{\prime }\left( z_{*}^{\left( 2\right) }\right) <0$ and $%
H_z^{\prime }\left( z_{*}^{\left( 1\right) }\right) >0$. Therefore, the
eigenvalues $E$ of (\ref{SchrEqu}) are bounded from below by the minimum
values of the potentials (\ref{UzPp}) and (\ref{UzPc}), and we denote $-%
\tilde \gamma $ and $\gamma $ the lowest eigenvalues of the respective
operators, as defined by (\ref{SchrEquPp}),(\ref{SchrEquPc}), with
eigenfunctions $\tilde \psi ^{\left( 1\right) }$ and $\psi ^{\left( 1\right)
}$. Exact solutions for $\tilde \psi ^{\left( 1\right) }$ and $\psi ^{\left(
1\right) }$ can be found for special cases of the inflaton potential $%
V\left( \varphi \right) $ (and special values of the parameters $\alpha $, $%
\beta $) for which the Schr\"odinger equation is exactly solvable (see
Appendix A).

Since the boundary conditions (\ref{zBC}) are homogeneous, the operator $%
\left( -1/2\right) \partial _{zz}+U\left( z\right) $ in (\ref{SchrEqu}) is
self-adjoint, and its eigenfunctions are orthogonal with respect to the
usual scalar product, 
\begin{equation}
\left\langle g_1\left( z\right) ,g_2\left( z\right) \right\rangle \equiv
\int g_1\left( z\right) g_2\left( z\right) dz.
\end{equation}
In the original variables, this scalar product becomes 
\begin{equation}
\left\langle P_1\left( \varphi \right) ,P_2\left( \varphi \right)
\right\rangle =\int P_1\left( \varphi \right) P_2\left( \varphi \right) 
\frac{\sqrt{2D\left( \varphi \right) }}{D\left( \varphi \right) ^\beta }\exp
\left( -\frac \pi {H\left( \varphi \right) ^2}\right) d\varphi .
\label{ScalProd}
\end{equation}

Using the scalar product (\ref{ScalProd}), the solution of the
time-dependent equation (\ref{TdFullEquPc}) with a given initial
distribution $P\left( \varphi ,t=0\right) \equiv P_0\left( \varphi \right) $
can be expressed through the orthonormal eigenfunctions $\psi ^{\left(
1\right) }$, $\psi ^{\left( 2\right) }$, $\psi ^{\left( 3\right) }$, ...
with eigenvalues $\gamma _{\left( 1\right) }\equiv -\gamma $, $\gamma
_{\left( 2\right) }$, $\gamma _{\left( 3\right) }$, ..., as 
\begin{equation}
P\left( \varphi ,t\right) =\sum_{n=1}^\infty C_n\exp \left( \gamma _{\left(
n\right) }t\right) \psi ^{\left( n\right) }\left( \varphi \right) ,
\label{PExpansion}
\end{equation}
with the coefficients $C_n$ given by 
\begin{equation}
C_n=\left\langle P_0\left( \varphi \right) ,\psi ^{(n)}\left( \varphi
\right) \right\rangle ,
\end{equation}
and similarly for $\tilde P\left( \varphi ,t\right) $. Note that co-moving
and physical volumes coincide at $t=0$, and thus $\tilde P\left( \varphi
,0\right) =P\left( \varphi ,0\right) $.

\subsection{The ratio of physical volumes, $r$}

The quantities of interest to us are the volumes ${\cal V}_{*}^{\ (j)}$ of
the thermalization hypersurfaces $\varphi =\varphi _{*}^{(j)}$. To express
these volumes in terms of the eigenfunctions (\ref{EigenFun}), we first
rewrite the diffusion equations (\ref{TdFullEquPp}),(\ref{TdFullEquPc}) as
continuity equations: 
\begin{eqnarray}
\partial _t\tilde P &=&-\partial _\varphi \tilde J+3H^\alpha \tilde P,
\label{ConsPp} \\
\partial _tP &=&-\partial _\varphi J.  \label{ConsPc}
\end{eqnarray}
Here, the fluxes $\tilde J$, $J$ are related to the distribution functions $%
\tilde P$, $P$ by 
\begin{equation}
J=-D\left( \varphi \right) ^{1/2+\beta }\partial _\varphi \left( D\left(
\varphi \right) ^{1/2-\beta }P\right) +v\left( \varphi \right) P.
\label{Flux}
\end{equation}
{}From the continuity equation (\ref{ConsPc}) it follows that the loss of
co-moving volume in inflating regions is compensated by the corresponding
growth in the volume of the thermalized regions. The rate of this growth is
given by the flux through the thermalization points: 
\begin{equation}
\frac \partial {\partial t}\int\limits_{\varphi _{*}^{\left( 1\right)
}}^{\varphi _{*}^{\left( 2\right) }}P\left( \varphi ,t\right) d\varphi
=\left. \left( -J\right) \right| _{\varphi _{*}^{\left( 1\right) }}^{\varphi
_{*}^{\left( 2\right) }}.
\end{equation}
Therefore, the co-moving volume ${\cal V}_{(c)*}^{\left( 1,2\right) }$
thermalized in a specific minimum of the inflaton potential throughout a
given time range is 
\begin{equation}
{\cal V}_{(c)*}^{\left( 1,2\right) }=\left| \int\limits_{t_1}^{t_2}J\left(
\varphi _{*}^{\left( 1,2\right) },t\right) dt\right|
\end{equation}
(the absolute value is needed to cancel the negative sign of the flux
through the leftmost thermalization point). We use a similar formula for the
thermalized physical volume: 
\begin{equation}
{\cal V}_{*}^{\left( 1,2\right) }=\left| \int\limits_{t_1}^{t_2}\tilde J%
\left( \varphi _{*}^{\left( 1,2\right) },t\right) dt\right|
\end{equation}

According to the method proposed in \cite{MakingPredictions}, the infinite
thermalization volumes 
\begin{equation}
{\cal V}_{*}^{\left( 1,2\right) }=\left| \int\limits_0^{t_\varepsilon
^{\left( 1,2\right) }}\tilde J\left( \varphi _{*}^{\left( 1,2\right)
},t\right) dt\right| =\left| \int\limits_0^{t_\varepsilon ^{\left(
1,2\right) }}v\left( \varphi _{*}^{\left( 1,2\right) }\right) \tilde P\left(
\varphi _{*}^{\left( 1,2\right) },t\right) dt\right|  \label{PhysVol}
\end{equation}
are to be cut off at times $t_\varepsilon ^{\left( 1,2\right) }$ which are
determined from the condition that only a fraction $\varepsilon $ of the
corresponding co-moving volume is left unthermalized at those times, 
\begin{equation}
\left| \int\limits_{t_\varepsilon ^{\left( 1,2\right) }}^\infty J\left(
\varphi _{*}^{\left( 1,2\right) },t\right) dt\right| =\varepsilon \left|
\int\limits_0^\infty J\left( \varphi _{*}^{\left( 1,2\right) },t\right)
dt\right| \equiv \varepsilon p_{1,2}.  \label{CoordVol}
\end{equation}
Here, $p_1$ and $p_2$ are fractions of the total co-moving volume (averaged
over an ensemble of universes) that will eventually thermalize in the first
and the second minima of the inflaton potential, respectively.

The integrals in (\ref{PhysVol}),(\ref{CoordVol}) are dominated by the
ground state eigenfunctions (\ref{EigenFun}) of (\ref{FullEquPp}),(\ref
{FullEquPc}) which we have denoted by $\tilde \psi ^{\left( 1\right) }\left(
\varphi \right) $ and $\psi ^{\left( 1\right) }\left( \varphi \right) $.
Using (\ref{PExpansion}) for $P\left( \varphi _{*}^{\left( 1,2\right)
},t\right) $, we obtain 
\begin{equation}
\int\limits_{t_\varepsilon ^{\left( 1,2\right) }}^\infty J\left( \varphi
_{*}^{\left( 1,2\right) },t\right) dt=\int\limits_{t_\varepsilon ^{\left(
1,2\right) }}^\infty v\left( \varphi _{*}^{\left( 1,2\right) }\right)
P\left( \varphi _{*}^{\left( 1,2\right) },t\right) dt\approx \frac{C_1\exp
\left( -\gamma t_\varepsilon ^{\left( 1,2\right) }\right) }\gamma v\left(
\varphi _{*}^{\left( 1,2\right) }\right) \psi ^{\left( 1\right) }\left(
\varphi _{*}^{\left( 1,2\right) }\right) .  \label{EpsEqu}
\end{equation}

After solving (\ref{CoordVol}),(\ref{EpsEqu})\ for $t_\varepsilon ^{\left(
1,2\right) }$, one can calculate from (\ref{PhysVol}) the ratio of the
physical volumes that thermalize in the two minima of the inflaton
potential: 
\begin{equation}
r\equiv \frac{{\cal V}_{*}^{\left( 2\right) }}{{\cal V}_{*}^{\left( 1\right)
}}=\frac{\left| v_2\right| \tilde \psi _2}{\left| v_1\right| \tilde \psi _1}%
\left( \frac{\left| v_2\right| \psi _2}{\left| v_1\right| \psi _1}\frac{p_1}{%
p_2}\right) ^{\tilde \gamma /\gamma },  \label{Ratio}
\end{equation}
where 
\begin{equation}
\psi _{1,2}\equiv \psi ^{\left( 1\right) }\left( \varphi _{*}^{\left(
1,2\right) }\right) ,\quad \tilde \psi _{1,2}\equiv \tilde \psi ^{\left(
1\right) }\left( \varphi _{*}^{\left( 1,2\right) }\right)  \label{Psi12}
\end{equation}
are the values of the ground state eigenfunctions taken at thermalization
points $\varphi _{*}^{\left( 1,2\right) }$, and 
\begin{equation}
v_{1,2}\equiv \left. -\frac{H^{\alpha -1}H^{\prime }}{4\pi }\right|
_{\varphi _{*}^{\left( 1,2\right) }}  \label{ConvVel}
\end{equation}
are ``drift velocities'' at those points.

\subsection{\label{Ambig-r}Ambiguities in $r$}

The form of the diffusion equations (\ref{TdFullEquPp}),(\ref{TdFullEquPc})
depends on the parameters $\alpha $ and $\beta $, and therefore the
quantities $\gamma $, $\tilde \gamma $, $\psi _j$, $\tilde \psi _j$, $p_j$
appearing in Eq. (\ref{Ratio}) for the volume ratio $r$ all depend on $%
\alpha $ and $\beta $. This dependence, which reflects the sensitivity of $r$
to the choice of the time parametrization and of the factor ordering, is the
main focus of the present paper and will be analyzed in detail in the
following two sections.

Another source of uncertainty is the choice of the initial distribution $%
P\left( \varphi ,0\right) \equiv P_0\left( \varphi \right) $ for the
calculation of $p_j$. If we choose a Gaussian distribution peaked at the
maximum of the potential $V\left( \varphi \right) $ and having a width $w$
much smaller than the characteristic width of the maximum, then the values
of $p_j$ should not be very sensitive to $w$. But a weak dependence of $p_j$
on $w$ certainly exists and should be addressed as a matter of principle.
Some numerical results on this $w$-dependence will be presented in Section 
\ref{NUMERIC}.

The problem of initial distribution can be resolved by invoking quantum
cosmology and using a distribution, like Eq. (\ref{TunWF}), obtained from
the cosmological wave function. However, it has been argued \cite{QCProb}
that probability is an approximate concept in quantum cosmology and can be
defined only within the semiclassical approximation. The accuracy of this
approximation for a nucleating universe is characterized by $%
S^{-1}=H_0^2/\pi $, where $S$ is the tunneling action and $H_0$ is the value
of $H$ at the maximum of $V\left( \varphi \right) $. Hence, it may be
impossible to reduce the uncertainty in $P_0\left( \varphi \right) $ below $%
O\left( H_0^2\right) $.

Yet another uncertainty in (\ref{Ratio}), this time rather benign, is
related to the already mentioned choice of the constant of order $1$ in (\ref
{ThermCond}) which affects the exact values of thermalization points $%
\varphi _{*}^{\left( 1,2\right) }$. A different choice will change the
behavior of the solution $\tilde \psi ^{\left( 1\right) }\left( \varphi
\right) $ near $\varphi =\varphi _{*}^{\left( 1,2\right) }$ (the derivative $%
\tilde \psi ^{\prime }\left( \varphi \right) $ changes sign at $\varphi
_{*}^{\left( 1,2\right) }$ \cite{LLMBig}). However, since the diffusion term
in (\ref{Flux}) is negligible at those points, one can employ the asymptotic
formulae \cite{MakingPredictions} 
\begin{equation}
\tilde \psi ^{\left( 1\right) }\left( \varphi \right) \approx \tilde c\frac{%
H\left( \varphi \right) }{H^{\prime }\left( \varphi \right) }\exp \left[
-12\pi \int_{\varphi _0}^\varphi \frac{H\left( \varphi \right) d\varphi }{%
H^{\prime }\left( \varphi \right) }+4\pi \tilde \gamma \int_{\varphi
_0}^\varphi \frac{H^{1-\alpha }\left( \varphi \right) d\varphi }{H^{\prime
}\left( \varphi \right) }\right] ,  \label{AsympPp}
\end{equation}
\begin{equation}
\psi ^{\left( 1\right) }\left( \varphi \right) \approx c\frac{H\left(
\varphi \right) }{H^{\prime }\left( \varphi \right) }\exp \left[ -4\pi
\gamma \int_{\varphi _0}^\varphi \frac{H^{1-\alpha }\left( \varphi \right)
d\varphi }{H^{\prime }\left( \varphi \right) }\right] ,  \label{AsympPc}
\end{equation}
which are accurate in the deterministic slow roll region up to the neglected
diffusion term. Here, $\varphi _0$ is some point in the region where
diffusion is negligible, 
\begin{equation}
H^2\left( \varphi _0\right) \ll \frac{\partial H}{\partial \varphi }\left(
\varphi _0\right) .  \label{NeglDiff}
\end{equation}
With the help of (\ref{AsympPp})-(\ref{AsympPc}), one can show that a change
in thermalization points $\varphi _{*}^{\left( 1,2\right) }\rightarrow \bar 
\varphi _{*}^{\left( 1,2\right) }$ changes the ratio (\ref{Ratio}) by the
ratio of the additional volume expansion factors gained before
thermalization, 
\begin{equation}
r\rightarrow r\frac{\exp \left( -12\pi \int_{\varphi _{*}^{\left( 2\right)
}}^{\bar \varphi _{*}^{\left( 2\right) }}\frac{H\left( \varphi \right)
d\varphi }{H^{\prime }\left( \varphi \right) }\right) }{\exp \left( -12\pi
\int_{\varphi _{*}^{\left( 1\right) }}^{\bar \varphi _{*}^{\left( 1\right) }}%
\frac{H\left( \varphi \right) d\varphi }{H^{\prime }\left( \varphi \right) }%
\right) }.  \label{AddInfl}
\end{equation}
To obtain unambiguous relative probabilities, one could compare the volumes
of constant-temperature hypersurfaces with the same value of $T$ in
different types of regions. Then one would have to multiply ${\cal V}%
_{*}^{(j)}$ by the additional expansion factors up to the chosen
temperature, and the dependence on the precise values of $\varphi
_{*}^{\left( 1,2\right) }$ would disappear.

\section{Analytic estimates}

\label{ANALYTIC}

\subsection{Estimate of eigenvalues}

\label{EstimateGammas}

Perhaps the most important parameters entering (\ref{Ratio}), in terms of
their effect on the magnitude of $r$, are the ground state eigenvalues of
Eqs. (\ref{FullEquPp}),(\ref{FullEquPc}). These eigenvalues $\tilde \gamma $%
, $\gamma $ can be estimated if the inflaton potential is sufficiently flat
and smooth near its maximum. The estimate is based on expansion of the
effective potential $U\left( z\right) $ of the Schr\"odinger equation (\ref
{SchrEqu}) around its minimum up to terms quadratic in $z$.

We shall assume for simplicity that the inflaton potential is symmetric
around its maximum at $\varphi =0$ up to terms of quartic order in $\varphi $%
. Then the expansion of $H\left( \varphi \right) $ around $\varphi =0$ has
the form 
\begin{equation}
H\left( \varphi \right) =H_0+\frac{H_2}2\varphi ^2+\frac{H_4}{24}\varphi
^4+O\left( \varphi ^5\right)  \label{SerPotential}
\end{equation}
with $H_0>0$ and $H_2<0$. The diffusion coefficient (\ref{DiffCoef}) is also
expanded as 
\begin{mathletters}
\begin{eqnarray}
D\left( \varphi \right) &=&D_0+\frac{D_2}2\varphi ^2+O\left( \varphi
^4\right) , \\
D_0 &=&\frac{H_0^{\alpha +2}}{8\pi ^2},\quad D_2=\frac{\left( \alpha
+2\right) H_0^{\alpha +1}H_2}{8\pi ^2}.
\end{eqnarray}
We then use (\ref{ChangeVar}) to find the derivatives with respect to $z$ at
the point $z=0$ corresponding to $\varphi =0$, 
\end{mathletters}
\begin{mathletters}
\begin{eqnarray}
\left. \frac{\partial H}{\partial z}\right| _{z=0} &=&0,\quad \left. \frac{%
\partial ^2H}{\partial z^2}\right| _{z=0}=2D_0H_2, \\
\left. \frac{\partial ^3H}{\partial z^3}\right| _{z=0} &=&0,\quad \left. 
\frac{\partial ^4H}{\partial z^4}\right| _{z=0}=8D_0D_2H_2+4D_0^2H_4,
\end{eqnarray}
and substitute them into (\ref{EffPotential}),(\ref{UzPp}) and (\ref{UzPc})
to obtain the expansion of the effective Schr\"odinger potential, 
\end{mathletters}
\begin{equation}
U\left( z\right) =U_0+\frac{\omega ^2}2z^2+O\left( z^3\right) .
\label{UQuadExp}
\end{equation}
The coefficients $U_0$ and $\omega $ for the potential (\ref{UzPc}) are
given by 
\begin{mathletters}
\begin{eqnarray}
U_0 &=&\left| H_2\right| \frac{H_0^{\alpha -1}}{8\pi }\left( 1-\beta H_0^2%
\frac{\alpha +2}{2\pi }\right) , \\
\omega ^2 &=&H_2^2\frac{H_0^{2\alpha -2}}{16\pi ^2}\left( 1-\frac{H_0^2}{%
2\pi }\left[ -5+2\alpha +2\beta \left( 2+\alpha \right) +\frac{H_4H_0}{H_2^2}%
\right] +O\left( H_0^4\right) \right)  \label{omegaSqPc}
\end{eqnarray}
and for the potential (\ref{UzPp}) 
\end{mathletters}
\begin{mathletters}
\begin{eqnarray}
\tilde U_0 &=&U_0-3H_0^\alpha =-3\,H_0^\alpha +\left| H_2\right| \frac{%
H_0^{\alpha -1}}{8\pi }\left( 1-\beta H_0^2\frac{\alpha +2}{2\pi }\right) ,
\\
\tilde \omega ^2 &=&H_2^2\frac{H_0^{2\alpha -2}}{16\pi ^2}\left( 1+12\alpha 
\frac{H_0^3}{\left| H_2\right| }-\frac{H_0^2}{2\pi }\left[ -5+2\alpha
+2\beta \left( 2+\alpha \right) +\frac{H_4H_0}{H_2^2}\right] +O\left(
H_0^4\right) \right) .  \label{omegaSqPp}
\end{eqnarray}
This analysis can be generalized to non-symmetric potentials with $H^{\prime
\prime \prime }\left( 0\right) \neq 0$; in that case, the minimum of $%
U\left( z\right) $ will be shifted from $z=0$.

Assuming that the quadratic expansion (\ref{UQuadExp}) of the potential $%
U\left( z\right) $ is accurate enough up to the classical turning points, we
can approximate the eigenvalues of (\ref{SchrEqu}) by the corresponding
eigenvalues of the harmonic oscillator, 
\end{mathletters}
\begin{equation}
E_n=U_0+\frac \omega 2+n\omega ,  \label{HarmEigenvalues}
\end{equation}
and the ground state eigenvalues by 
\begin{eqnarray}
-\tilde \gamma &=&\tilde U_0+\frac{\tilde \omega }2,  \label{GpEst0} \\
\gamma &=&U_0+\frac \omega 2.  \label{GcEst0}
\end{eqnarray}
Although our boundary conditions are not the same as those for the harmonic
oscillator eigenvalues (\ref{HarmEigenvalues}), they are imposed at points $%
\varphi _{*}^{(1,2)}$ deeply within the classically forbidden region of the
Schr\"odinger equation and it seems reasonable to assume that a different
choice of these boundary conditions does not significantly alter the
eigenvalues. The validity of the estimates (\ref{GpEst0})-(\ref{GcEst0}) was
confirmed numerically (see Section \ref{NUMERIC}).

We assume that the potential $V\left( \varphi \right) $ is flat near its
maximum and that the maximum value $V\left( 0\right) $ is small in Planck
units: 
\begin{equation}
\frac{\left| H_2\right| }{H_0}=\left. \frac{\left| V^{\prime \prime }\right| 
}{2V}\right| _{\varphi =0}\ll 1,\quad H_0^2=\frac{8\pi V\left( 0\right) }3%
\ll 1.  \label{H2}
\end{equation}
We also assume that 
\begin{equation}
\frac{H_4H_0}{H_2^2}=-3+2\left. \frac{V^{\left( \text{IV}\right) }V}{\left(
V^{\prime \prime }\right) ^2}\right| _{\varphi =0}\lesssim 1,  \label{H4}
\end{equation}
which is, for instance, the case for the potential 
\begin{equation}
V\left( \varphi \right) =\lambda \left( \varphi ^2-\eta ^2\right) ^2.
\label{ExamplePotential}
\end{equation}
Under these assumptions, and if the values of $\alpha $ and $\beta $ are not
unreasonably large, the estimates (\ref{GpEst0})-(\ref{GcEst0}) give: 
\begin{equation}
\tilde \gamma \approx 3\,H_0^\alpha \left[ 1-\frac{\left| H_2\right| }{H_0}%
\frac 1{24\pi }\left( 1+\sqrt{1+12\alpha \frac{H_0^3}{\left| H_2\right| }}%
\right) +O\left( H_0^2\right) \right] ,  \label{GampEst}
\end{equation}
and 
\begin{equation}
\gamma \approx \frac{\left| H_2\right| }{H_0}\frac{H_0^\alpha }{4\pi }\left(
1-\frac{H_0^2}{8\pi }\left[ -5+2\alpha +4\beta \left( 2+\alpha \right) +%
\frac{H_4H_0}{H_2^2}+O\left( H_0^2\right) \right] \right) .  \label{GamcEst}
\end{equation}
The ratio of the ground state eigenvalues is therefore estimated as 
\begin{equation}
\frac{\tilde \gamma }\gamma \approx 12\pi \frac{H_0}{\left| H_2\right| }%
\left[ 1-\frac 1{24\pi }\frac{\left| H_2\right| }{H_0}\left( 1+\sqrt{%
1+12\alpha \frac{H_0^3}{\left| H_2\right| }}\right) +O\left( H_0^2\right)
\right] .  \label{GamRatio}
\end{equation}
Here, the dependence on $\alpha $ and $\beta $ has been absorbed in $O\left(
H_0^2\right) $, except for the $\alpha $-dependent term under the square
root. Note that it follows from (\ref{H2}),(\ref{GamRatio}) that $\tilde 
\gamma /\gamma \gg 1$.

To evaluate the $\alpha $-dependence of the ratio (\ref{GamRatio}) resulting
from the square root term, we have to consider the magnitude of $%
12H_0^3/\left| H_2\right| $. If 
\begin{equation}
12\frac{H_0^3}{\left| H_2\right| }\ll 1,  \label{Cond1}
\end{equation}
then the square root in (\ref{GamRatio}) can be expanded in powers of $%
H_0^3/\left| H_2\right| $, which gives 
\begin{equation}
\frac{\tilde \gamma }\gamma \approx 12\pi \frac{H_0}{\left| H_2\right| }%
\left[ 1-\frac{\left| H_2\right| }{12\pi H_0}+\frac{H_0^2}{16\pi }\left(
-5+6\beta \left( 2+\alpha \right) +\frac{H_4H_0}{H_2^2}\right) +\dots
\right].  \label{GamRatio1}
\end{equation}
Here we have explicitly written the $O\left( H_0^2\right) $ terms, and the
ellipsis represents higher-order terms. In the opposite limit, 
\begin{equation}
12\frac{H_0^3}{\left| H_2\right| }\gg 1,  \label{Cond2}
\end{equation}
the $\alpha $-dependent term in square brackets in (\ref{GamRatio}) becomes $%
\left( \alpha H_0\left| H_2\right| /48\pi ^2\right) ^{1/2}\ll H_0^2\sqrt{%
\alpha }$, and we obtain 
\begin{equation}
\frac{\tilde \gamma }\gamma \approx 12\pi \frac{H_0}{\left| H_2\right| }%
\left[ 1-\frac{\left| H_2\right| }{24\pi H_0}+\frac{H_0^2}{8\pi }\left(
-5+2\alpha +4\beta \left( 2+\alpha \right) +\frac{H_4H_0}{H_2^2}\right)
+\dots \right] .  \label{GamRatio2}
\end{equation}
We see that, in both cases, the dependence on $\alpha $ and $\beta $ appears
only in terms of order $O\left( H_0^2\right) $ and higher.

\subsection{Accuracy and limits of applicability}

The estimates (\ref{GampEst})-(\ref{GamRatio}) for the eigenvalues of the
diffusion equation are valid if the inflaton potential is sufficiently
smooth and flat near its maximum, so that the effective potential of the
Schr\"odinger equation (\ref{SchrEqu}) could be approximated as in (\ref
{UQuadExp}) within a sufficiently large region including the classical
turning point of (\ref{SchrEqu}). This holds if the $z^4$ term in the
expansion of $U\left( z\right) $, 
\begin{equation}
U\left( z\right) =U_0+\frac{\omega ^2}2z^2+\frac{U_4}{24}z^4,\quad U_4=\frac{%
5\left| H_2\right| ^3H_0^{3\alpha -1}}{16\pi ^4}\left( 1-\frac{2\alpha }5-%
\frac{H_4H_0}{5H_2^2}+O\left( H_0^2\right) \right),
\end{equation}
is smaller than the quadratic term, $\left( \omega ^2/2\right) z^2$, at the
classical turning point $z_0$ found from 
\begin{equation}
U_0+\frac{\omega ^2}2z_0^2=U_0+\frac \omega 2,\quad z_0=\frac 1{\sqrt{\omega 
}},
\end{equation}
which gives the condition 
\begin{equation}
\frac{5H_0^2}{6\pi }\left| 1-\frac{2\alpha }5-\frac{H_4H_0}{5H_2^2}\right|
\ll 1.
\end{equation}
This inequality holds as long as conditions (\ref{H2}) and (\ref{H4}) are
satisfied and $\alpha $ is not very large. One can show that the analogous
condition for $\tilde U\left( z\right) $ holds as well.

Another assumption we made in deriving (\ref{GpEst0})-(\ref{GcEst0}) was $%
\omega ^2>0$, $\tilde \omega ^2>0$, meaning that the potentials $U\left(
z\right) $ and $\tilde U\left( z\right) $ have a minimum (and not a maximum)
at $z=0$. If (\ref{H2})-(\ref{H4}) hold and $\alpha $ and $\beta $ are of
order $1$, then $\omega ^2>0$ also holds, and the condition $\tilde \omega
^2>0$ becomes 
\begin{equation}
1+12\alpha \frac{H_0^3}{\left| H_2\right| }>0,\quad \alpha >-\frac{\left|
H_2\right| }{12H_0^3}.
\end{equation}
This condition may be violated for large negative $\alpha $, but it holds
for $\left| \alpha \right| \sim 1$ if the condition (\ref{Cond1}) is
satisfied.

Now we consider the accuracy of the expression (\ref{Ratio}) for the volume
ratio $r$. Since it contains the ratio $\tilde \gamma /\gamma $ in the
exponent, and the estimate (\ref{GamRatio}) for $\tilde \gamma /\gamma $
gives an ambiguity of order $O\left( H_0^2\right) $, the result of (\ref
{Ratio}) is generally reliable only with logarithmic precision, i.e. $\ln r$
is determined with a relative accuracy of order $O\left( H_0^2\right) $.
However, if (\ref{Cond1}) is satisfied, the $\alpha $- and $\beta $%
-dependent terms in (\ref{GamRatio1}) will be much smaller than $1$, and\
the result of (\ref{Ratio}) itself will be accurate up to $O\left(
H_0^3/\left| H_2\right| \right) $. In this limiting case, we are able to
derive a simpler approximate formula for the volume ratio $r$. We notice
that under the condition (\ref{Cond1}), the region where the quadratic
expansion (\ref{UQuadExp}) of the potential $U\left( z\right) $ is valid, 
\begin{equation}
\frac{\omega ^2}2z^2\gg \frac{U_4}{24}z^4,\ \left| z\right| \ll \frac{\omega 
\sqrt{12}}{\sqrt{U_4}},\ \left| \varphi \right| \ll \sqrt{\frac{H_0}{\left|
H_2\right| }},  \label{HOReg}
\end{equation}
overlaps with the region where diffusion is negligible, 
\begin{equation}
H^2\ll \left| \frac{\partial H}{\partial \varphi }\right| \approx \left|
H_2\varphi \right| ,\ \left| \varphi \right| \gg \varphi _0=\frac{H_0^2}{%
\left| H_2\right| }.  \label{NoDiffReg}
\end{equation}
Then, the ground state eigenfunction $\psi _{HO}\left( \varphi \right) $ of
the harmonic oscillator potential (\ref{UQuadExp}), being a good
approximation to the eigenfunction of (\ref{SchrEqu}) in the region (\ref
{HOReg}), can be matched with the asymptotic solution in the no-diffusion
region, 
\begin{equation}
\psi ^{(j)}\left( \varphi \right) =\frac{c^{(j)}}{v\left( \varphi \right) }%
\exp \left[ -4\pi \gamma \int_{\varphi _0^{(j)}}^\varphi \frac{H^{1-\alpha }%
}{H^{\prime }}d\varphi \right] .
\end{equation}
Since the match points $\varphi _0^{(j)}$ lie within the region (\ref{HOReg}%
), and the solution $\psi _{HO}\left( \varphi \right) $ is symmetric in that
region, this gives $c^{(1)}=c^{(2)}$. Analogous results are obtained for the
physical volume distribution, $\tilde \psi ^{(j)}\left( \varphi \right) $.
The formula (\ref{Ratio}) for the volume ratio $r$, written in terms of
coefficients $c^{(j)}$, $\tilde c^{(j)}$, becomes \cite{MakingPredictions} 
\begin{equation}
r=\frac{\tilde c^{(2)}}{\tilde c^{(1)}}\left( \frac{c^{(2)}}{c^{(1)}}\frac{%
p_1}{p_2}\right) ^{\tilde \gamma /\gamma }\left( \frac{Z_{*}^{\left(
1\right) }}{Z_{*}^{\left( 2\right) }}\right) ^3.  \label{RatioC}
\end{equation}
With the additional assumption which we verified numerically, 
\begin{equation}
\frac{p_1}{p_2}=1+O\left( H_0^2\right) ,  \label{PRatio}
\end{equation}
Eq. (\ref{RatioC}) becomes 
\begin{equation}
r=\left( \frac{Z_{*}^{\left( 1\right) }}{Z_{*}^{\left( 2\right) }}\right)
^3\left( 1+O\left( \frac{H_0^3}{\left| H_2\right| }\right) \right) ,
\label{SimpleRatio}
\end{equation}
where 
\begin{equation}
Z_{*}^{\left( 1,2\right) }=\exp \left[ -4\pi \int_{\pm \varphi _0}^{\varphi
_{*}^{(1,2)}}\frac{H^{\prime }}Hd\varphi \right]  \label{ZFactor}
\end{equation}
are the volume expansion factors during deterministic slow roll. The value
of $\varphi _0$ is unimportant as long as it lies within the region (\ref
{HOReg}) where the potential is symmetric.

\section{Numerical results}

\label{NUMERIC} To check our analytic estimates of the eigenvalues (\ref
{GampEst}),(\ref{GamcEst}) and to find the dependence of the volume ratio (%
\ref{Ratio}) on the parameters $\alpha $ and $\beta $, we performed
numerical calculations. The inflaton potential was chosen as 
\begin{equation}
V\left( \varphi \right) =V_0\left( 1-\frac{\varphi ^2}{\eta ^2}+\kappa
\left( \frac \varphi \eta \right) ^5\exp \left( -\mu \frac{\varphi ^2}{\eta
^2}\right) \right) ^2,  \label{OurPotential}
\end{equation}
where dimensionless parameters $\kappa $ and $\mu $ characterize the
asymmetry of the potential. We considered both the symmetric case, $\kappa
=0 $, and the non-symmetric case $(\kappa ,\mu \neq 0)$. The calculation of
the volume ratio (\ref{Ratio}) was performed in the non-symmetric case,
since it is identically equal to $1$ for a symmetric potential.

The conditions (\ref{H2})-(\ref{H4}) for the potential (\ref{OurPotential})
are satisfied if 
\begin{equation}
V_0\ll 1,\quad \eta ^2\gg 1.
\end{equation}

\subsection{Technique}

To find the eigenvalues of the stationary equations (\ref{SchrEquPp}),(\ref
{SchrEquPc}), we used the standard 4th-5th order Runge-Kutta method \cite
{NumRecipes}. To facilitate the solution when $P\left( \varphi ,t\right) $
varies greatly in order of magnitude, we rewrote the stationary versions of
the Eqs. (\ref{ConsPp})-(\ref{Flux}) in the dimensionless variables $\ln
P\left( \varphi ,t\right) $ and $\left( J/P\right) $ as follows: 
\begin{eqnarray}
\partial _\varphi \left( \ln P\right) &=&-8\pi ^2\left( \frac JP\right)
-\left( 2\pi +\left( \frac \alpha 2+1\right) H^2\right) H^{\prime }H^{\alpha
-3}, \\
\partial _\varphi \left( \frac JP\right) &=&\gamma -\left( \frac JP\right)
\partial _\varphi \left( \ln P\right) , \\
\partial _\varphi \left( \ln \tilde P\right) &=&-8\pi ^2\left( \frac{\tilde J%
}{\tilde P}\right) -\left( 2\pi +\left( \frac \alpha 2+1\right) H^2\right)
H^{\prime }H^{\alpha -3}, \\
\partial _\varphi \left( \frac{\tilde J}{\tilde P}\right) &=&3H^\alpha -%
\tilde \gamma -\left( \frac{\tilde J}{\tilde P}\right) \partial _\varphi
\left( \ln \tilde P\right) ,
\end{eqnarray}
and solved for the smallest values of $\gamma $ and $-\tilde \gamma $ to
satisfy the boundary conditions, 
\begin{equation}
\left. \frac JP\right| _{\varphi _{*}^{\left( 1,2\right) }}=v_{1,2},
\end{equation}
where $v_{1,2}$ is given by (\ref{ConvVel}). The resulting eigenfunctions $%
\tilde \psi ^{\left( 1\right) }$, $\psi ^{\left( 1\right) }$ were used to
calculate the values $\tilde \psi _{1,2}$, $\psi _{1,2}$ of (\ref{Psi12}).

For the numerical solution of the time-dependent equations (\ref{TdFullEquPp}%
) and (\ref{TdFullEquPc}), we have used the (slightly modified)
unconditionally stable Crank-Nicholson finite difference scheme \cite
{NumRecipes} with boundary conditions (\ref{NoDiffBC}) and the initial
distribution given by a Gaussian, $P\left( \varphi ,t=0\right) \propto \exp
\left( -\varphi ^2/w^2\right) $, with the width $w\ll \eta $. The solution $%
P\left( \varphi ,t\right) $ was used to obtain the values $p_{1,2}$ of (\ref
{CoordVol}).

\subsection{Symmetric potentials}

The numerical calculation of the eigenvalues $\tilde \gamma $, $\gamma $ for
symmetric potentials (\ref{OurPotential}) was performed to verify the
estimates (\ref{GampEst})-(\ref{GamcEst}). The numerically obtained
eigenvalues $\tilde \gamma $, $\gamma $ and deviations from the estimates $%
\tilde \gamma _0$, $\gamma _0$ for $H_0=\sqrt{8\pi V_0/3}=.05$ and $\eta =8$
as well as the ratio $\tilde \gamma /\gamma $ are summarized in the
following tables. The eigenvalues were found with relative precision of $%
10^{-7}$.

\begin{center}
\begin{table}[tbph]

\begin{tabular}{|c|r|r|r|r|r|}
$\alpha $ & -1.0 & -0.5 & 0.0 & 0.5 & 1.0 \\ \hline
$\gamma $ & 0.04982 & 0.01114 & 0.00249 & 5.57e-04 & 1.2445e-04 \\ \hline
$\left( \gamma -\gamma _0\right) /\gamma _0$ & 4.94e-04 & 9.71e-05 & 2.99e-04
& 6.9e-04 & 8.31e-04 \\ \hline
$\tilde \gamma $ & 59.9694 & 13.4067 & 2.9975 & 0.6702 & 0.14985 \\ \hline
$\left( \tilde \gamma -\tilde \gamma _0\right) /\tilde \gamma _0$ & 7.52e-06
& 5.87e-06 & 3.05e-07 & 2.14e-07 & 2.e-07 \\ \hline
$\tilde \gamma /\gamma $ & 1203.60 & 1203.59 & 1203.70 & 1203.83 & 1203.98
\end{tabular}
\caption{Eigenvalues for $\beta=-1$.} 
\end{table}

\begin{table}[tbph]

\begin{tabular}{|c|r|r|r|r|r|}
$\alpha $ & -1.0 & -0.5 & 0.0 & 0.5 & 1.0 \\ \hline
$\gamma $ & 0.04980 & 0.01113 & 0.002489 & 5.565e-04 & 1.244e-04 \\ \hline
$\left( \gamma -\gamma _0\right) /\gamma _0$ & 6.94e-04 & 5.94e-04 & 4.95e-04
& 3.97e-04 & 2.97e-04 \\ \hline
$\tilde \gamma $ & 59.9694 & 13.4068 & 2.99751 & 0.6702 & 0.149851 \\ \hline
$\left( \tilde \gamma -\tilde \gamma _0\right) /\tilde \gamma _0$ & 6.69e-06
& 2.12e-07 & 7.e-16 & 1.e-16 & 2.e-07 \\ \hline
$\tilde \gamma /\gamma $ & 1204.08 & 1204.07 & 1204.17 & 1204.31 & 1204.45
\end{tabular}
\caption{Eigenvalues for $\beta =0$.} 
\end{table}

\begin{table}[tbph]

\begin{tabular}{|c|r|r|r|r|r|}
$\alpha $ & -1.0 & -0.5 & 0.0 & 0.5 & 1.0 \\ \hline
$\gamma $ & 0.04978 & 0.01113 & 0.002488 & 5.5628e-04 & 1.243e-04 \\ \hline
$\left( \gamma -\gamma _0\right) /\gamma _0$ & 8.93e-04 & 1.09e-03 & 1.29e-03
& 1.49e-03 & 1.69e-03 \\ \hline
$\tilde \gamma $ & 59.9695 & 13.4068 & 2.997 & 0.6702 & 0.149851 \\ \hline
$\left( \tilde \gamma -\tilde \gamma _0\right) /\tilde \gamma _0$ & 5.85e-06
& 2.e-07 & 3.52e-07 & 3.95e-07 & 4.17e-07 \\ \hline
$\tilde \gamma /\gamma $ & 1204.56 & 1204.55 & 1204.65 & 1204.78 & 1204.93
\end{tabular}
\caption{Eigenvalues for $\beta =1$.} 
\end{table}
\end{center}

While the eigenvalues themselves vary significantly with $\alpha $, the
ratio $\tilde \gamma /\gamma $ is very nearly constant. One can see that the
variance in the eigenvalue ratio $\tilde \gamma /\gamma $ due to changes in
factor ordering parameter $\beta $ is comparable to the variance due to
changes in the time variable parameter $\alpha $. We have performed
numerical calculations using other values of $H_0$ and $\eta $ in the ranges 
$H_0=0.5$---$0.001$ and $\eta =1$---$20$ and obtained similar results.

\subsection{Asymmetric potentials}

Here we present our numerical results for the potential (\ref{OurPotential}).

For $H_0=.05$, $\eta =5$, $\kappa =20$, and $\mu =50$, the eigenvalue ratios
and the values of the volume ratio (\ref{Ratio}) are summarized in the
following tables.

\begin{center}
\begin{table}[tbph]

\begin{tabular}{|r|r|r|r|}
$\alpha $ 
\mbox{$\backslash$}
$\beta $ & -1.0 & 0.0 & 1.0 \\ \hline
-1.0 & 469.5152 & 469.7020 & 469.8889 \\ \hline
0.0 & 469.5978 & 469.7846 & 469.9716 \\ \hline
1.0 & 469.6986 & 469.8855 & 470.0726
\end{tabular}
\caption{Eigenvalue ratio $\tilde \gamma / \gamma$.\label{AsymmGamRatio}}%
\end{table}

\begin{table}[tbp] \centering

\begin{tabular}{|r|r|r|r|}
$\alpha $ 
\mbox{$\backslash$}
$\beta $ & -1.0 & 0.0 & 1.0 \\ \hline
-1.0 & 279.1466 & 279.2707 & 279.3949 \\ \hline
0.0 & 278.1129 & 278.2344 & 278.3560 \\ \hline
1.0 & 277.6672 & 277.7888 & 277.9106
\end{tabular}
\caption{Logarithm of the volume ratio (\ref{Ratio}).\label{AsymmVolRatio}}%
\end{table}
\end{center}

As the tables show, the relative change in the eigenvalue ratio is $\sim
10^{-3}$, which agrees with the estimate $O\left( H_0^2\right) \approx
2\cdot 10^{-3}$. Also, it is clear that the dependence on $\alpha $ is of
the same order as the dependence on $\beta $. Calculations were performed
for other values of the parameters with the same conclusions.

The form (\ref{OurPotential}) of the inflaton potential was chosen to allow
for analytic estimates of section \ref{EstimateGammas}, namely, the
assumption that the third derivative $H^{\prime \prime \prime}\left(
0\right) $ vanishes holds for this potential. We have performed numerical
calculations for a potential with nonvanishing $H^{\prime \prime
\prime}\left( 0\right) $ and obtained similar results.

To verify the assumption (\ref{PRatio}) used in our derivation of (\ref
{SimpleRatio}), we looked at the dependence of $\ln \left( p_1/p_2\right) $
on $\alpha $ and $\beta $. Our results suggest that the relative variance of 
$\ln \left( p_1/p_2\right) $ with $\alpha $ and $\beta $ is also of the
order $H_0^2$. However, in our case of a ``flat top'' potential, the
symmetry of the potential in the diffusive region leads to $p_1 \approx p_2$%
, so that 
\begin{equation}
\frac{\tilde \gamma }\gamma \ln \frac{p_1}{p_2}\ll 1,
\end{equation}
and the volume ratio (\ref{SimpleRatio}) is virtually unaffected by the
dependence of $\ln \left( p_1/p_2\right) $ on $\alpha $ and $\beta $. For
potentials with asymmetry in the diffusive region, the value of $\ln \frac{%
p_1}{p_2}$ was of the order $1$, however its dependence on $\alpha $ and $%
\beta $ remained small ($O(H_0^2)$).

Another possible source of uncertainty discussed above was the choice of the
initial distribution. We performed calculations of the volume ratio (\ref
{Ratio}) for a Gaussian initial distribution, 
\[
P_0\left( \varphi \right) \propto \exp \left[ -\left( \frac \varphi {w\eta }%
\right) ^2\right] 
\]
with varying width parameters $w=0.0001$---$0.01$, and the results varied
insignificantly (the relative change in $\ln r$ was of the order $10^{-5}$).

\section{Conclusions}

\label{DISCUSSION}

In this paper we analyzed the ambiguities in assigning probabilities to
different types of thermalized volumes in an eternally inflating universe.
One of the key factors determining the relative probabilities is the volume
ratio $r$, given by Eq. (\ref{Ratio}). Our results are most easily
formulated in terms of the quantity $y=\ln r$.

We introduced a parameter $\alpha $ representing the ambiguity due to the
choice of the time variable $t$, and parameter $\beta $ representing the
factor ordering ambiguity in the diffusion equation (\ref{TdFullEquPp}). Our
main result, obtained both analytically and numerically, is that variation
of either of these parameters introduces a variation 
\begin{equation}
\delta y/y\sim V_0,  \label{Accuracy}
\end{equation}
where $V_0\sim H_0^2$ is the highest value of the inflaton potential $%
V\left( \varphi \right) $, and $H_0$ is the expansion rate of the universe
at $V=V_0$.

Since the factor ordering ambiguity is inherent in the diffusion
approximation, Eq. (\ref{Accuracy}) gives a bound on the accuracy of the
predictions of the model. Moreover, since variations of $y$ arising from
varying $\alpha $ and $\beta $ are of the same order, we conclude that
within that accuracy, the results are independent of the choice of time
variable. This is an intriguing result, since it suggests that in a more
fundamental approach, based e.g. on the Wheeler-De Witt equation,
probabilities could be manifestly independent of time parametrization.

Another source of uncertainty is the choice of the initial distribution for
the calculation of $p_j$ in Eq.\ (\ref{Ratio}). Our results indicate that
the corresponding variation of $y$ is even smaller than (\ref{Accuracy}).
One could try to avoid this uncertainty by using an initial distribution,
like Eq.\ (\ref{TunWF}), derived from the cosmological wave function.
However, probability is an approximate concept in quantum cosmology, and
unitarity holds only with the accuracy of the semiclassical approximation 
\cite{QCProb}. For a nucleating universe, this accuracy is of order $V_0$
(see Sec. \ref{Ambig-r}).

Throughout the paper we have assumed inflation of the ``new'' type with a
potential well below the Planck scale, $V\left( \varphi \right) \ll 1$. In
this case, $\delta y/y\ll 1$. As the maximum value of the potential $V_0$ is
increased, the accuracy gradually deteriorates, and errors become $O\left(
1\right) $ when it reaches the Planck scale, $V_0\sim 1$. This is expected
to happen in models of ``chaotic'' inflation, where the probability
distribution $\tilde P\left( \varphi ,t\right) $ is concentrated near the
Planck values of the potential \cite{Footnote0,Footnote1}.

The relative uncertainty in the volume ratio $r$ itself is typically greater
than (\ref{Accuracy}), due to the presence of the large exponent, $\tilde 
\gamma /\gamma \sim V_0/\left| V_0^{\prime \prime }\right| $, in Eq. (\ref
{Ratio}): 
\begin{equation}
\frac{\delta r}r=\delta y\sim \frac{V_0^2}{\left| V_0^{\prime \prime
}\right| }.  \label{Accuracy-r}
\end{equation}
This is small if $V_0^2/\left| V_0^{\prime \prime }\right| \ll 1$, in which
case $r$ is accurately given by the simple formula (\ref{SimpleRatio}).
Otherwise, the uncertainty in $r$ is large. We note, however, that the
probabilities of thermalization into different minima of $V\left( \varphi
\right) $ are expected to be vastly different, so that the corresponding
values of $y$ are large, $\left| y\right| \gg 1$, and the resulting volume
ratios are either very large ($r>>>1$) or very small ($r<<<1$). These strong
inequalities are unaffected by the uncertainty (\ref{Accuracy}). It can
affect only rare borderline cases when the two probabilities are nearly
equal. It appears that we have to accept this as a genuine uncertainty of
the problem and make predictions only in cases where one minimum is much
more probable than the other.

In this paper, we considered exclusively the problem of finding the relative
probabilities for thermalization into different minima of the inflaton
potential in a single universe \cite{GBNew}. The same method can also be
applied \cite{Predictions} to an ensemble of disconnected, eternally
inflating universes with different potentials $V\left( \varphi \right) $,
parametrized by some variable ``constants of Nature'' $\left\{ \lambda
_j\right\} $.\ (The set of allowed values of $\left\{ \lambda _j\right\} $
may be either discrete or continuous.) The volume of thermalized regions in
a given universe would then depend on the cutoff parameter $\epsilon $ as 
\begin{equation}
{\cal V}_{*}\propto \epsilon ^{-\tilde \gamma /\gamma },
\end{equation}
where the eigenvalues $\tilde \gamma $, $\gamma $ pertain to the diffusion
equation with the potential $V\left( \varphi \right) $ in that universe. In
the limit $\epsilon \rightarrow 0$, only universes with maximum value of $%
\tilde \gamma /\gamma $ will have a non-zero probability \cite{Footnote2}.

It is possible that the condition 
\begin{equation}
\tilde \gamma /\gamma =\max  \label{MaxGamRatio}
\end{equation}
selects a unique set of $\left\{ \lambda _j\right\} $. Then the potential $%
V\left( \varphi \right) $ is fixed, and the only remaining problem is to
find the relative probabilities for thermalization into different minima of
this potential. On the other hand, it is conceivable that the maximum of $%
\tilde \gamma /\gamma $ is strongly degenerate \cite{Footnote3}, so that Eq.
(\ref{MaxGamRatio}) selects a large subset of all allowed values of $\left\{
\lambda _j\right\} $. It is worth noting that the class of potentials having
the same ``ground state'' eigenvalue $\gamma $ is very wide: it can be
parametrized by an arbitrary function (see Appendix B). The relative
probabilities for $\left\{ \lambda _j\right\} $ within the degenerate subset
can be calculated following the same procedure as in Ref. \cite
{MakingPredictions} and in the present paper.

\section*{Notes added}

1. After this paper was submitted, we learned about an interesting paper 
\cite{MezhNew}, in which arbitrary time parametrizations (\ref{GenTime}) for
the diffusion equation (\ref{TdFullEquPp}) were considered. The authors
described a transformation of potential $V\left( \varphi \right) $ and a
corresponding change of time variables that give identical physical
predictions at late times. In the framework of the present paper, that
equivalence transformation leaves the Schr\"odinger equation (\ref{SchrEqu})
and boundary conditions (\ref{zBC}) unchanged, giving the same eigenvalues
and eigenfunctions and, therefore, identical predictions. The family (\ref
{SolvablePotentials}) of exactly solvable potentials was also found in \cite
{MezhNew}.

2. In a recent preprint \cite{AlternReg}, Linde and Mezhlumian suggested a
family of alternative regularization procedures parameterized by a
dimensionless number $q$. All these procedures have the same property of
time reparametrization invariance as the one we discussed here, indicating
that the invariance requirement alone is not sufficient to select a unique
regularization procedure. In this note we shall briefly discuss some
additional requirements which may fix the parameter $q$.

The alternative regularizations of \cite{AlternReg} are identical to ours,
except the co-moving volume distribution $P\left( \varphi ,t\right) $ is
replaced by a weighted distribution $P_q\left( \varphi ,t\right) $ which
satisfies a modified version of Eq. (\ref{ConsPp}), 
\begin{equation}
\partial _tP_q+\partial _\varphi J_q=3q\frac H{T\left( H\right) }P_q.
\label{NewConsPq}
\end{equation}
(The value $q=0$ corresponds to the unmodified regularization procedure of 
\cite{MakingPredictions}.) For $q>0$, a greater weight is assigned to
co-moving regions which have expanded by a greater factor; for $q<0$, to
regions which have expanded by a smaller factor. The latter situation
appears somewhat unnatural, and we can require that 
\begin{equation}
q\geq 0.  \label{QPos}
\end{equation}

The parameter $q$ should be chosen so that the smallest eigenvalue $\gamma _q
$ of (\ref{NewConsPq}) satisfies 
\begin{equation}
\gamma _q>0,  \label{GamQCond}
\end{equation}
ensuring that the integrals in Eq. (\ref{CoordVol}) are convergent. Using
the scale factor time parametrization ($\alpha =0$) we have $\gamma
_q=\gamma -3q$, where $\gamma \equiv \gamma _{q=0}$ takes values in the
range $0<\gamma <3$, depending on the inflaton potential $V\left( \varphi
\right) $. If we require that the regularization scheme should apply to all
possible inflaton potentials, then the condition (\ref{GamQCond}) restricts $%
q$ to the range $q\leq 0$. Combining this with the condition (\ref{QPos}),
we are left with a single value, $q=0$.

Let us finally consider the situation when the potential $V\left( \varphi
\right) $ in Fig. 1 is symmetric in the range of $\varphi $ where diffusion
is non-negligible, so that the difference between ${\cal V}_{*}^{(1)}$ and $%
{\cal V}_{*}^{(2)}$ is due entirely to the regions of deterministic slow
roll. We can require that in this case the volume ratio should be the same
as in the case of non-stochastic, finite inflation, $r=\left(
Z_{*}^{(2)}/Z_{*}^{(1)}\right) ^3$, where $Z_{*}^{(1)}$ and $Z_{*}^{(2)}$
are the corresponding slow-roll expansion factors. Once again, this selects $%
q=0$.

Although none of the conditions we have suggested appears to be mandatory,
the above discussion does suggest that the regularization scheme with $q=0$
has some unique features and may therefore be preferred.

\section*{Acknowledgments}

The authors are grateful to A. Linde for insightful discussions and to A.
Mezhlumian for his comments and for informing us about the paper \cite
{MezhNew}. This work was supported in part by NSF.

\appendix

\section{Exact solutions of the diffusion equation}

For specific inflaton potentials $V\left( \varphi \right) $, the
Schr\"odinger equation (\ref{SchrEqu}) is exactly solvable. We will present
several examples of such potentials.

The widest range of solvable potentials occurs with the choice $\alpha
=\beta =0$. In this case, equations (\ref{SchrEquPp}) and (\ref{SchrEquPc})
differ only by a constant shift of the eigenvalue, and it is sufficient to
consider one equation (\ref{SchrEquPc}). The potential (\ref{UzPc}) is given
by 
\begin{equation}
U\left( z\right) =\frac 12b^{\prime }\left( z\right) +\frac 12b^2\left(
z\right) ,
\end{equation}
where 
\begin{equation}
b\left( z\right) =-\pi \frac{H^{\prime }\left( z\right) }{H^3\left( z\right) 
}.
\end{equation}
The simplest case of a solvable Schr\"odinger equation is that with a
constant (or piecewise-constant) potential, $U\left( z\right) =U_0$. This is
achieved, for instance, if 
\begin{equation}
b\left( z\right) =\pm \sqrt{2U_0},\quad H\left( z\right) =\left[ \pm \frac{2%
\sqrt{2U_0}}\pi \left( z-z_0\right) \right] ^{-1/2},\quad z=z_0\pm 2\pi 
\sqrt{2U_0}\varphi ^2,
\end{equation}
which corresponds to 
\begin{mathletters}
\begin{eqnarray}
V\left( \varphi \right) &=&\frac 3{64U_0\varphi ^2},\quad U_0\neq 0,
\label{Unz} \\
V\left( \varphi \right) &=&const,\quad U_0=0.  \label{Uzero}
\end{eqnarray}
Taking $V\left( \varphi \right) $ to be of the form (\ref{Unz}) and (\ref
{Uzero}) with different $U_0$ on different regions of values of $\varphi $,
one obtains a piecewise-constant potential $U\left( z\right) =U_0$ for the
Schr\"odinger equation (\ref{SchrEquPc}).

Another exactly solvable case is that of the harmonic oscillator potential, 
\end{mathletters}
\begin{equation}
U\left( z\right) =U_0+\frac{\omega ^2}2z^2,
\end{equation}
with 
\begin{equation}
U_0=\pm \frac \omega 2,\quad b\left( z\right) =\pm \omega z,\quad H\left(
z\right) =\left( \pm \frac{\omega z^2}\pi +c\right) ^{-1/2},
\end{equation}
The upper choice of sign corresponds to the inflaton potential

\begin{equation}
V\left( \varphi \right) =V_0\frac{\exp \left( -4\sqrt{\pi \omega }\varphi
\right) }{\left( 1+c\exp \left( -4\sqrt{\pi \omega }\varphi \right) \right)
^2}.  \label{SolvablePotentials}
\end{equation}
Particular cases of these potentials include the exponential potential \cite
{Salopek}, $V\left( \varphi \right) =V_0\exp \left( a\varphi \right) $, and
the potentials 
\begin{equation}
V\left( \varphi \right) =\frac{V_0}{\cosh ^2a\varphi },\quad V\left( \varphi
\right) =\frac{V_0}{\cos ^2a\varphi }.
\end{equation}

\section{Potentials with a given ground state eigenvalue}

It is always possible to choose a potential for a Schr\"odinger equation
that would possess a given ground state eigenvalue, and the freedom of this
choice is parametrized by an arbitrary function. One can show this using
methods of supersymmetric quantum mechanics \cite{SUSYQM}. We start with an
arbitrary function $f\left( z\right) $ and define the potential $U\left(
z;f,E_0\right) $ for the Schr\"odinger equation (\ref{SchrEqu}) by 
\begin{equation}
U\left( z;f,E_0\right) \equiv \frac 12\exp \left( -f\left( z\right) \right) 
\frac{d^2}{dz^2}\exp f\left( z\right) +E_0=\frac 12\left( f^{\prime \prime
}+f^{\prime 2}\right) +E_0.  \label{PotE0}
\end{equation}
It is easily verified that the function $g\left( z\right) =\exp f\left(
z\right) $ satisfies (\ref{SchrEqu}) with eigenvalue $E_0$. Since the
function $g\left( z\right) $ is everywhere nonzero, it would be the ground
state eigenfunction if appropriate boundary conditions were satisfied.
Generic homogeneous boundary conditions, 
\begin{equation}
\left. \left( g^{\prime }+B\left( z\right) g\right) \right| _{z_j}=0,
\end{equation}
imposed at some boundary points $z_{1,2}$, translate into boundary
conditions imposed on $f\left( z\right) $: 
\begin{equation}
f^{\prime }\left( z_j\right) =-B\left( z_j\right) .
\end{equation}
Since the boundary conditions restrict the behavior of $f\left( z\right) $
only near the boundary points, the freedom of choosing a function $f\left(
z\right) $ is essentially unaffected by the boundary condition requirement.
We see that for all functions $f\left( z\right) $ the corresponding
potential $U\left( z;f,E_0\right) $ has the required ground state eigenvalue 
$E_0$.

It is also always possible to find an inflaton potential $V\left( \varphi
\right) $ for Eq. (\ref{TdFullEquPc}) that will lead to $U\left(
z;f,E_0\right) $ in the corresponding Schr\"odinger equation. To do that,
one needs to find a solution $g_0\left( z\right) $ of the equation (\ref
{SchrEqu}) with potential (\ref{PotE0}) with eigenvalue $0$ (where the
function $g_0\left( z\right) $ does not have to satisfy the boundary
conditions). Then one can define $b\left( z\right) \equiv g_0^{\prime }/g_0$
and express the potential $U\left( z;f,E_0\right) $ as in (\ref{UzPc}), 
\begin{equation}
U\left( z;f,E_0\right) =\frac 12\left( b^{\prime }+b^2\right) .
\end{equation}
Using the terminology of SUSY quantum mechanics, the function $-b\left(
z\right) $ is the ``superpotential''. By transforming back to the physical
variable $\varphi $, one can obtain the potential $V\left( \varphi \right) $
corresponding to this Schr\"odinger equation.


\begin{references}
\bibitem{Constants}  More exactly, the probabilities are proportional to the
volumes of the thermalization hypersurfaces which are the boundaries of the
corresponding thermalized regions. For details, see Refs. \cite
{MakingPredictions,Predictions}.
\bibitem{AmbigPhi}  The behavior of the derivative $\tilde \psi ^{\prime
}\left( \varphi \right) $ is significantly modified near $\varphi
_{*}^{\left( 1,2\right) }$. It has been shown in \cite{LLMBig} that $\tilde 
\psi \left( \varphi \right) $ develops very shallow minima near these
points, so that $\tilde \psi ^{\prime }\left( \varphi \right) $ changes
sign. However, the magnitude of the function $\tilde \psi \left( \varphi
\right) $ itself is essentially unaffected by this modification.
\bibitem{Method}  This is achieved by a generalization of the method used in 
\cite{Japanese} for the potential $V\left( \varphi \right) =\lambda \varphi
^4$ and in {\ \cite{Mijic}} for a particular factor ordering and time
variable. The transformation into a self-adjoint form can similarly be
performed for a general factor ordering (\ref{GenFactor}) and time variable (%
\ref{GenTime}).
\bibitem{Footnote0}  A related observation was made by Linde and Mezhlumian 
\cite{LLMBig} who noted sensitive dependence of the results on the boundary
condition they imposed at the Planck scale.
\bibitem{Footnote1}  This is not to say that no predictions at all are
possible in models with $V_0\sim 1$. Consider for example a model with a
potential as in Fig. 1, where instead of thermalization, the points $\varphi
_{*}^{(1,2)}$ correspond to the onset of the second period of inflation,
driven by another inflaton field, $\chi $. (This is the situation in models
of ``hybrid'' inflation \cite{HybridInflation}.) Suppose that the potentials 
$V^{\left( 1,2\right) }\left( \chi \right) $ are also in the form of a
double well, but now with maxima well below the Planck scale. So, altogether
we have 4 minima in which the vacuum energy can thermalize. Since $V_0\sim 1$%
, we cannot accurately predict whether the universe is more likely to end up
in the pair of minima with $\varphi >0$ or with $\varphi <0$. However, we
should be able to calculate the relative probability for each pair. Another
example is the calculation of the probability distribution for density
fluctuations on various scales (see Ref. \cite{MakingPredictions}). In both
cases the probabilities can be accurately calculated because they are
unaffected by what happens at $V\left( \varphi \right) \sim 1$.
\bibitem{Footnote2}  It is easy to see that our scheme gives nonsensical
results if all possible forms of the potential $V\left( \varphi \right) $
are allowed. Then the maximum of $\tilde \gamma /\gamma $ would be achieved
for $V\left( \varphi \right) =const$, which gives $\gamma =0,$ $\tilde \gamma
=3,$ and $\tilde \gamma /\gamma =\infty $. Nontrivial results are obtained
only if the set of allowed potentials is sufficiently restricted.
\bibitem{Footnote3}  Note also that if we accept Eq. (\ref{Accuracy}) as the
genuine uncertainty of the model, then values of $\tilde \gamma /\gamma $
with a relative deviation of order $V_0$ from the maximum should still be
assigned nonzero probabilities.
\end{references}
\end{document}